\documentclass[aip,jcp]{revtex4-1}

\usepackage{graphicx}

\newcommand{\beq}{\begin{equation}}
\newcommand{\eeq}{\end{equation}}
\newcommand{\bea}{\begin{eqnarray}}
\newcommand{\eea}{\end{eqnarray}}

\begin{document}





\title{Inhibition of DNA ejection from bacteriophage by Mg$^{+2}$ counterions}








\author{SeIl Lee,$^1$ C. V. Tran,$^2$ and T. T. Nguyen$^1$}




\affiliation{$^1$School of Physics, Georgia Institute of Technology, 
837 State Street, Atlanta, Georgia 30332-0430\\
$^2$School of Chemistry and Biochemistry, Georgia Institute of Technology, 
901 Atlantic Drive, Atlanta, Georgia 30332-0400}





\date{\today}

\begin{abstract}

The problem of inhibiting viral DNA ejection from bacteriophages by
multivalent counterions, specifically Mg$^{+2}$ counterions,
is studied. Experimentally, it is known that MgSO$_4$ 
salt has a strong and non-monotonic effect on the amount of DNA ejected. There
exists an optimal concentration at which the minimum amount of DNA is ejected 
from the virus. At lower or higher concentrations, more DNA is ejected from the capsid. 
We propose that this phenomenon is the result of DNA overcharging by Mg$^{+2}$ 
multivalent counterions. As Mg$^{+2}$ concentration increases from zero, the
net charge of DNA 
changes from negative to positive. The optimal inhibition corresponds to the
 Mg$^{+2}$ concentration where DNA is neutral. At lower/higher concentrations, 
DNA genome is charged. It prefers to be in solution to lower its electrostatic
self-energy, which consequently leads to an increase in DNA ejection. 
By fitting our theory to available experimental data, the strength of 
DNA$-$DNA short range attraction energies, mediated by Mg$^{+2}$, 
is found to be $-$0.004 $k_BT$ per nucleotide base. 
This and other fitted parameters agree well with known values from
other experiments and computer simulations. The parameters are
also in aggreement qualitatively with values for tri- and tetra-valent
counterions.

\end{abstract}


\maketitle 















%




%











%






\section{Introduction}

Most bacteriophages, or viruses that infect bacteria, are composed of 
a DNA genome coiling inside a rigid, protective capsid. 
It is well-known that the persistence
length, $l_p$, of DNA is about 50 nm, comparable to or larger than the inner 
diameter of the viral capsid. 
The genome of a typical bacteriophage is about 10 microns
or 200 persistence lengths. Thus the DNA molecule is 
considerably bent and strongly confined inside the viral capsid
resulting in a substantially pressurized capsid with 
internal pressure as high as 
50 atm [\onlinecite{Bustamante01,Gelbart03,Gelbart2003,Harvey07}].
It has been suggested that this
pressure is the main driving force for the ejection of the
viral genome
into the host cell when the capsid tail binds to
the receptor in the cell membrane, and subsequently opens the capsid.
This idea is supported by various experiments both
\textit{in vivo} and \textit{in vitro}
[\onlinecite{Santamaria04,Gelbart03,Black89,Murialdo91,Gelbart2003,Phillips05, Gelbart04,Knobler08}].
The \textit{in vitro} experiments additionally revealed possibilities of
controlling
the ejection of DNA from bacteriophages. One example is the addition of
PEG (polyethyleneglycol), a large molecule incapable of penetrating
the viral capsid. A finite PEG concentration in solution
produces an apparent osmotic pressure on the capsid.
This in turn leads to a reduction
or even complete inhibition of the ejection of DNA.

\begin{figure}[ht]
\resizebox{9cm}{!}{\includegraphics{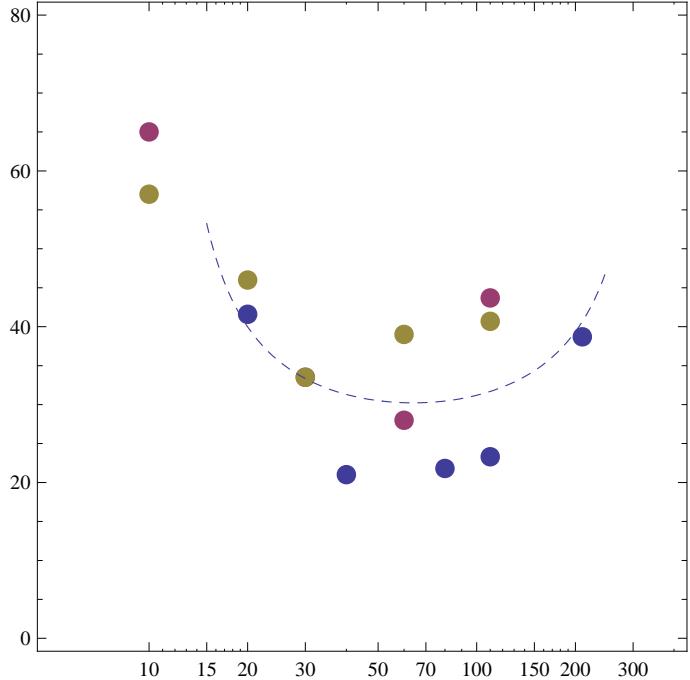}}
\caption{(Color online) Inhibition of DNA ejection depends on MgSO$_4$ concentration
for bacteriophage $\lambda$ at 3.5 atm external osmotic pressure.
Solid circles represent experimental data from Ref. 
[\onlinecite{Knobler08, GelbartPriCom09}],
where different colors corresponds to different experimental
batch. The dashed line is a theoretical fit of our theory.
See Sec. \ref{sec:fitting}
}
\label{fig:Mg2}
\end{figure}

Since DNA is a strongly charged molecule in aqueous solution, the screening
condition of the solution also affects the ejection process.
At a given external osmotic pressure,
by varying the salinity of solution, one can also vary the amount
of DNA ejected. Interestingly, it has been shown that monovalent
counterions such as NaCl have a negligible effect on 
the DNA ejection process [\onlinecite{Gelbart03}].
In contrast, multivalent counterions such as Mg$^{+2}$, 
CoHex$^{+3}$ (Co-hexamine),
Spd$^{+3}$ (spermidine) or Spm$^{+4}$ (spermine) exert
strong effect. 
In this paper, we focus on the role of Mg$^{+2}$ divalent
counterion on DNA ejection. In Fig. \ref{fig:Mg2},
the percentage of ejected DNA from bacteriophage $\lambda$ 
(at 3.5 atm external osmotic pressure) from
the experiment of Ref. \onlinecite{Knobler08,GelbartPriCom09} are plotted
as a function of MgSO$_4$ concentration (solid circles). 
The three colors correspond to three different sets of data.
Evidently, the effect of multivalent counterions
on the DNA ejection is non-monotonic. There is an optimal Mg$^{+2}$
concentration where the minimum amount of DNA genome 
is ejected from the phages.
%

The general problem of understanding 
DNA condensation and interaction in the presence of multivalent
counterions is rather complex, as evident by the large literature
dedicated to this subject. This is especially true in the case
of divalent counterions because many physical
factors involved are energetically comparable to each other. 
Most studies related to DNA screening in the presence of divalent
counterions have focused on ion specific effects. For example,
in Ref. \onlinecite{Knobler08},
hydration effects were proposed to explain the data of
DNA ejection in the presence of MgCl$_2$ salt where 
the minimum has not yet been observed for salt concentration
upto 100 mM. In this paper, we focus on understanding the 
non-specific electrostatic interactions involved in the inhibition of DNA 
ejection by divalent counterions. We show that some aspects of the
DNA ejection experiments can be explained within this
framework. Specifically, we propose that the non-monotonic 
behavior observed in Fig. 
\ref{fig:Mg2} has similar physical origin to that of the phenomenon
of the reentrant condensation of macroions
in the presence of multivalent counterions.
It is the result of Mg$^{+2}$ ions inducing an 
effective attraction between DNA segments inside the capsid, and
the so-called overcharging of DNA by multivalent counterions
in free solution. 

Specifically, the electrostatics of Mg$^{+2}$ modulated
DNA ejection from bacteriophages is following. 
Due to strong electrostatic interaction
between DNA and Mg$^{+2}$ counterions, the counterions condense
on the DNA molecule. As a result, a DNA molecule behaves electrostatically
as a charged polymer with the effective net charge,
$\eta^*$ per unit length, equal to the sum of the ``bare" DNA charges, 
$\eta_0=-1e/1.7$\AA, and the charges of condensed 
counterions. There are strong correlations between 
the condensed counterions at the DNA surface which cannot be 
described using the standard 
Poisson-Boltzmann mean-field theory. Strongly correlated counterion
theories, 
various experiments and simulations 
[\onlinecite{Shklovskii1999,NguyenRMP2002,Netz02,lemayNature2007,KanduJCP2010}] 
have showed that 
when these strong correlations are taken into account,
$\eta^*$ is not only smaller than $\eta_0$ in magnitude but
can even have opposite sign: this is known as the 
\textit{charge inversion} phenomenon.
The degree of counterion condensation, and correspoly the value of
$\eta^*$, depends logarithmically on the concentration of 
multivalent counterions, $N_Z$. As $N_Z$ increases from zero, 
$\eta^*$ becomes less negative, 
neutral and eventually positive. We propose that the 
multivalent counterion concentration, $N_{Z,0}$,
where DNA's net charge is neutral corresponds to the optimal inhibition
due to Mg$^{+2}-$induced
DNA-DNA attraction inside the capsid. At counterion concentration
$N_Z$ lower or higher than $N_{Z,0}$, $\eta^*$ is either negative or positive.
As a charged molecule at these concentrations, DNA prefers to be in 
solution to lower its electrostatic self-energy (due to the geometry
involved, the capacitance of DNA molecule is higher in free solution 
than in the bundle inside the capsid). Accordingly, this leads to a 
higher percentage of ejected viral genome. 

The fact that Mg$^{+2}$ counterions can have such strong influence on DNA ejection
is highly non-trivial. It is well-known that Mg$^{+2}$ ions do not condense or 
only condense partially free DNA molecules in aqueous 
solution [\onlinecite{Parsegian92}, \onlinecite{Hud01}]. Yet, they exert strong
effects on DNA ejection from bacteriophages.
We argue that this is due to the entropic confinement of the viral
capsid. Unlike free DNA molecules in solution, 
DNA packaged inside capsid are strongly bent and 
the thermal fluctuations of
DNA molecule is strongly suppressed.
It is due to this unique setup of the bacteriophage
where DNA is pre-packaged by a motor protein
during virus assembly that Mg$^{+2}$ ions can induce attractions
between DNA. It should be mentioned that Mg$^{+2}$ counterions 
have been shown {\em experimentally} to
condense DNA in another confined system:
the DNA condensation in two dimension [\onlinecite{Koltover2000}]. 
Recent computer simulations [\onlinecite{NguyenPRL2010}, \onlinecite{Nordenskiold95}]
 also show that 
if the lateral motion of DNA is restricted, divalent counterions can 
induced DNA condensation. The strength of DNA$-$DNA attraction energy
mediated by divalent counterions is comparable
to the results presented in this paper. These facts strongly
support our proposed argument.

The dashed line in Fig. \ref{fig:Mg2} is a fit of our theoretical 
result to the experimental data for MgSO$_4$. The optimal Mg$^{+2}$ concentration
is shown to be $N_{Z,0}=64$ mM. 
The Mg$^{+2}-$mediated attraction between DNA double helices 
is found to be $-0.004$ $k_BT$/base ($k_B$ is the Boltzmann constant and
$T$ is the temperature of the system). As discussed later
in Sec. \ref{sec:fitting}, these values agree well with various known
parameters of other DNA systems.

The organization of the paper as follows; In Sec. 
\ref{sec:overcharging}, a brief review of
the phenomenon of overcharging DNA by multivalent counterions
is presented. In Sec. III, the semi-empirically theory is fit to
the experimental data of DNA ejection from bacteriophages.
In Sec. \ref{sec:fitting}, the obtained fitting parameters is discussed
in the context of various other experimental and simulation studies of
DNA condensation by divalent counterions. 
Finally, we conclude our paper in Sec. V.

\section{Overcharging of DNA by multivalent counterions}

\label{sec:overcharging}

In this section, let us briefly visit
the phenomenal of overcharging of DNA by multivalent counterions 
to introduce various physical parameters involved in our theory. 
Standard linearized mean field theories of electrolyte solution states that
in solutions with mobile ions, the Coulomb potential of a point 
charge, $q$, is screened exponentially beyond a 
Debye-H\"{u}ckel (DH) screening radius, $r_s$:
\beq
V_{DH}(r) = \frac{q}{r} \exp (-r/r_s) .
\eeq
The DH screening radius $r_s$ depends on the 
concentrations of mobile ions in solution and is given by:
\beq
r_s = \sqrt{\frac{D k_BT}{4\pi e^2\sum_i c_i z_i^2} }
\eeq
where $c_i$ and $z_i$ are the concentration and the valence of mobile
ions of species $i$, $e$ is the charge of a proton,
and $D\approx 78$ is the dielectric constant of water.

Because DNA is a strongly charged molecule in solution, linear
approximation breaks down near the DNA surface because the
potential energy, $eV_{DH}(r)$, would be greater than $k_BT$ 
in this region. It has been shown that, within the 
general non-linear meanfield Poisson-Boltzmann theory, 
the counterions would condense on the DNA surface
to reduce its surface potential to be about $k_BT$. This
so-called Manning counterion condensation effect
leads to an ``effective" DNA linear charge density:
\beq
\eta_c = - Dk_BT /e
\eeq

In these mean field theories, the charge
of a DNA remains negative at all ranges of ionic strength of the solution.
The situation is completely different when DNA is screened by
multivalent counterions such as Mg$^{2+}$, Spd$^{3+}$ or Spm$^{4+}$.
These counterions also condense on DNA surface due to theirs strong
attraction to DNA negative surface charges. 
However, unlike their monovalent counterparts, the electrostatic
interactions among condensed counterions are very strong due
to their high valency. These interactions are even stronger than $k_BT$ and
mean field approximation is no longer valid in this case.
Counterintuitive phenomena emerge when DNA molecules
are screened by multivalent counterions. For example, 
beyond a threshold counterion concentration, 
the multivalent counterions can even over-condense on 
a DNA molecule making its net charge {\em positive}. Furthermore,
near the threshold concentration, DNA molecules are neutral
and they can attract each other causing condensation of DNA
into macroscopic bundles (the so-called {\em like-charged attraction}
phenomenon). 

\begin{figure}[ht]
\resizebox{9cm}{!}{\includegraphics{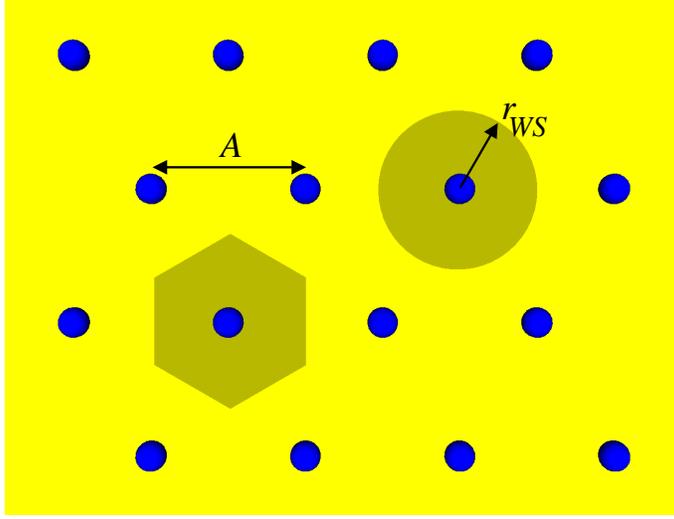}}
\caption{(Color online) Strong electrostatic interactions among condensed 
counterions lead to the formation of a strongly correlated
liquid on the surface of the DNA molecule. In the limit of 
very strong interaction, this liquid forms a two-dimensional
Wigner crystal with lattice constant $A$. 
The shaded hexagon is a Wigner-Seitz
cell of the background charge. It can be approximated as
a disc of radius $r_{WS}$.
}
\label{fig:WC}
\end{figure}

To understand how multivalent counterions overcharge DNA molecules,
let us write down the balance of the electro-chemical potentials
of a counterion at the DNA surface and in the bulk solution.
\beq
\mu_{cor} + Ze \phi(a)+ k_BT \ln [N_Z(a)v_o] = k_BT \ln [N_Z v_o].
\label{muBalanceEq}
\eeq
Here $v_o$ is the molecular volume of the counterion, $Z$ is 
the counterion valency. $\phi(a)$ is the
electrostatic surface potential at the dressed DNA. Approximating the 
dressed DNA as a uniformly charged cylinder with linear charged density
$\eta^*$ and radius $a$, $\phi(a)$ can be written as:
\beq
\phi(a) = \frac{2\eta^*}{D} \frac{K_0(a/r_s)}{(a/r_s) K_1(a/r_s)}
\simeq \frac{2\eta^*}{D} \ln \big(1+\frac{r_s}{a}\big)
\label{phi02}
\eeq
where $K_0$ and $K_1$ are Bessel functions (this expression is twice 
the value given in Ref. \onlinecite{Winterhalter88} because
we assume that the screening ion atmosphere does not penetrate the
DNA cylinder). 
In Eq. (\ref{muBalanceEq}), $N_Z(a)$ is the local concentration of the counterion at 
the DNA surface:
\beq
N_Z(a) \approx \sigma_0 / (Ze\lambda ) = \eta_0 / (2\pi a Ze \lambda)
\eeq
where $\sigma_0 = \eta_0 / 2\pi a$ is the bare surface charge
density of a DNA molecule and
the Gouy-Chapman length $\lambda=Dk_BT/2\pi\sigma_0Ze$ is 
the distance at which the potential energy of a counterion due 
to the DNA bare surface charge is one thermal energy $k_BT$.
The term $\mu_{cor}$ in Eq. (\ref{muBalanceEq}) is due to the
correlation energies of the counterions at the DNA surface.
It is this term which is {\em neglected} in mean-field theories. 
Several approximate, 
complementary theories, such as strongly correlated liquid
[\onlinecite{Shklovskii1999, Shklovskii99, NguyenRMP2002}], 
strong coupling [\onlinecite{Netz02}, \onlinecite{KanduJCP2010}] or counterion 
release [\onlinecite{BruinsmaCounterionRelease1998}, \onlinecite{GelbartPhysToday}]
have been proposed to 
calculate this term. Although with varying degree of analytical
complexity, they have similar physical origins. 
In this paper, we followed the theory presented in Ref. \onlinecite{NguyenRMP2002}.
In this theory, the strongly interacting counterions in the
condensed layer are assumed to form a two-dimensional strongly 
correlated liquid on the surface of the DNA (see Fig. \ref{fig:WC}).
In the limit of very strong correlation, the liquid form a two-dimensional
Wigner crystal (with lattice constant $A$) and $\mu_{cor}$ is proportional 
to the interaction energy of the counterion with
background charges of its Wigner-Seitz cell. Exact calculation
of this limit gives [\onlinecite{NguyenRMP2002}]:
\beq
\mu_{cor} \approx -1.65\frac{(Ze)^2}{Dr_{WS}} = -1.17 \frac{1}{D}~ (Ze)^{3/2} 
      \left(\frac{\eta_0}{a}\right)^{1/2} .
\label{eq:mucor}
\eeq
Here $r_{WS}=\sqrt{\sqrt{3}A^2/2\pi}$ is the radius of a disc with the same area as that
of a Wigner-Seitz cell of the Wigner crystal (see. Fig. \ref{fig:WC}).
It is easy to show that for multivalent counterions, 
the so-called Coulomb coupling (or plasma) parameter, 
$\Gamma = (Ze)^2/Dr_{WS}k_BT$, is greater than one. Therefore,
$|\mu_{cor}| > k_BT$, and thus cannot be neglected
in the balance of chemical potential, Eq. (\ref{muBalanceEq}).

Knowing $\mu_{cor}$, one can easily solve Eq. (\ref{muBalanceEq})
to obtain the net charge of a DNA for a given counterion
concentration:
\beq
\eta^* = -\frac{Dk_BT}{2Ze}\frac{\ln(N_{Z,0}/N_Z)}{\ln (1+r_s/a)},
\label{Eq:charge}
\eeq
where the concentration $N_{Z,0}$ is given by:
\beq
N_{Z,0} = N_Z(a) e^{-|\mu_{cor}|/k_BT}
\label{Eq:c0analytical}
\eeq

Eq. (\ref{Eq:charge}) clearly shows that 
for counterion concentrations higher than $N_{Z,0}$,
the DNA net charge $\eta^*$ is positive, indicating the 
over$-$condensation of the counterions on DNA. 
In other words, DNA is overcharged by multivalent 
counterions at these concentrations. Notice  
Eq. (\ref{eq:mucor}) shows that, 
for multivalent counterions $Z\gg 1$, $\mu_{cor}$ is strongly 
negative for multivalent counterions,  $|\mu_{cor}| \gg k_BT$.
Therefore, $N_{Z,0}$ is exponentially smaller than $N_Z(a)$
and a realistic concentration obtainable in experiments.

\begin{figure}[ht]
\resizebox{15cm}{!}{\includegraphics{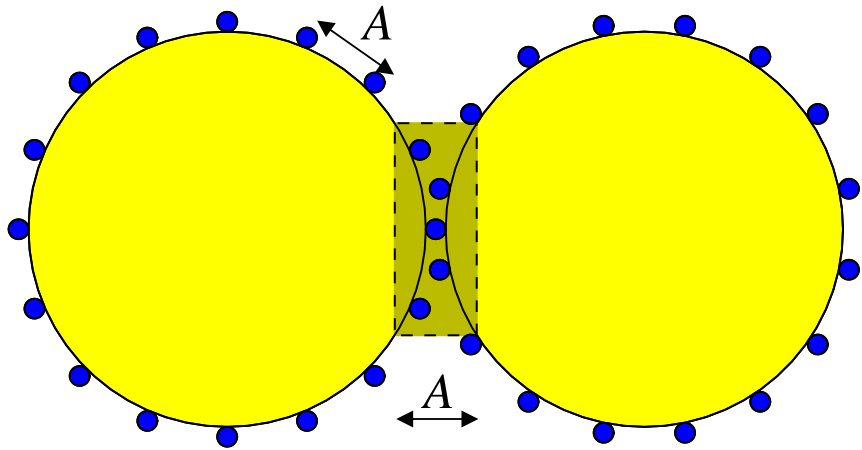}}
\caption{(Color online) Cross section of two touching DNA
molecules (large yellow circles) with condensed counterions (blue circles).
At the place where DNA touches each other (the
shaded region of width $A$ shown),
the density of the condensed counterion layer doubles and
additional correlation energy is gained. This leads to a 
short range attraction between the DNA molecules.
}
\label{fig:2DNA}
\end{figure}

Besides the overcharging phenomenon, DNA molecules
screened by multivalent counterions also experience 
the counterintuitive like-charge attraction effect. 
This short range attraction between DNA molecules 
can also be explained within the framework of the strong correlated liquid
theory. Indeed, in the area where DNA molecules touch each
other, each counterion charge is compensated
by the "bare'' background charge of {\em two} DNA molecules instead
of one (see Fig. \ref{fig:2DNA}). 
Due to this doubling of background charge, each counterion condensed in this region
gains an energy of:
\beq
\delta \mu_{cor} \approx \mu_{cor}(2\eta_0) - \mu_{cor}(\eta_0) 
  \simeq -0.46 \frac{1}{D}~ (Ze)^{3/2} \left(\frac{\eta_0}{a}\right)^{1/2} .
\eeq 
As a result, DNA molecules experience a short range correlation-induced 
attraction. Approximating the width
of this region to be on the order of the Wigner crystal
lattice constant $A$, the DNA$-$DNA attraction per unit length can
be calculated:
\beq
\mu_{\mbox{DNA}} \simeq - \frac{2\sqrt{2aA}\sigma_0}{Ze} |\delta\mu_{cor}|
\simeq -0.34 \frac{1}{D} \eta_0^{5/4} \left(\frac{Ze}{a}\right)^{3/4}
\label{Eq:muAnalytical}
\eeq

The combination of the overcharging of DNA molecules
and the like charged attraction
phenomena (both induced by multivalent counterions) leads to
the so-called reentrant condensation of DNA. At small counterion
concentrations, $N_Z$, DNA molecules are undercharged. At
high counterion concentrations, $N_Z$, DNA molecules are overcharged.
The Coulomb repulsion between charged DNA molecules 
keeps individual DNA molecules apart in solution.
At an intermediate range of $N_Z$, DNA molecules are mostly
neutral. The short range attraction forces are able to
overcome weak Coulomb repulsion leading to their condensation.
In this paper, we proposed that this reentrant behavior
of DNA condensation as function of counterion concentration
is the main physical mechanism behind the non-monotonic
dependence of DNA ejection from bacteriophages as
a function of the Mg$^{+2}$ concentrations.


\section{Theoretical calculation of DNA ejection from bacteriophage}

We are now in the position to obtain a theoretical description
of the problem of DNA ejection from bacteriophages in the
presence of multivalent counterions. We begin by writing the total 
energy of a viral DNA molecule as the sum 
of the energy of DNA segments ejected outside the capsid
with length $L_o$ and the energy of DNA segments 
remaining inside the capsid with
length $L_i = L-L_o$, where $L$ is the total length of the viral DNA genome: 
%
\beq
E_{tot}(L_o) = E_{in}(L_i) + E_{out}(L_o)
\label{Eq:Etotal}
\eeq

Because the ejected DNA segment is under no entropic confinement,
we neglect contributions from bending energy and 
approximate $E_{out}$ by the electrostatic
energy of a free DNA of the same length in solution:
\beq
E_{out}(L_o) = -L_o ({{\eta^*}^2}/{D}) \ln (1+{r_s}/{a}),
\label{Eq:Eout}
\eeq
where the DNA net charge, $\eta^*$, for a given
counterion concentration is given by Eq. (\ref{Eq:charge}).
The negative sign in Eq. (\ref{Eq:Eout}) signifies the 
fact that the system of the combined DNA and the condensed counterions
is equivalent to a cylindrical capacitor under 
constant charging potential.
As shown in previous section, we expect the $\eta^*$ to be
a function of the multivalent counterion concentration $N_Z$ 
and can be positive when $N_Z > N_{Z,0}$. 
In the limit of strongly correlated liquid, 
$N_{Z,0}$ is given in Eq. (\ref{Eq:c0analytical}). 
However, the exponential factor in
this equation shows that an accurate evaluation of $N_{Z,0}$
is very sensitive to an accurate calculation of the correlation
chemical potential $\mu_{cor}$. For practical purposes,
the accurate calculation of $\mu_{cor}$ is a highly
non-trivial task. One would need to go beyond the flat two-dimensional 
Wigner crystal approximation and takes into account not only
the non-zero thickness of the condensed counterion layer
but also the complexity of DNA geometry. 
Therefore, within the scope of this paper, 
we are going to consider $N_{Z,0}$ as a phenomenological 
constant concentration whose value is obtained by fitting 
the result of our theory to the experimental data.

The energy of the DNA segment inside the viral capsid comes
from the bending energy of the DNA coil and the interaction
between neighboring DNA double helices:
\beq
E_{in}(L_i,d) = E_{bend}(L_i,d) + E_{int}(L_i,d).
\label{Eq:Ein}
\eeq
where $d$ is the average DNA$-$DNA interaxial distance.
\begin{figure}[ht]
\resizebox{8cm}{!}{\includegraphics{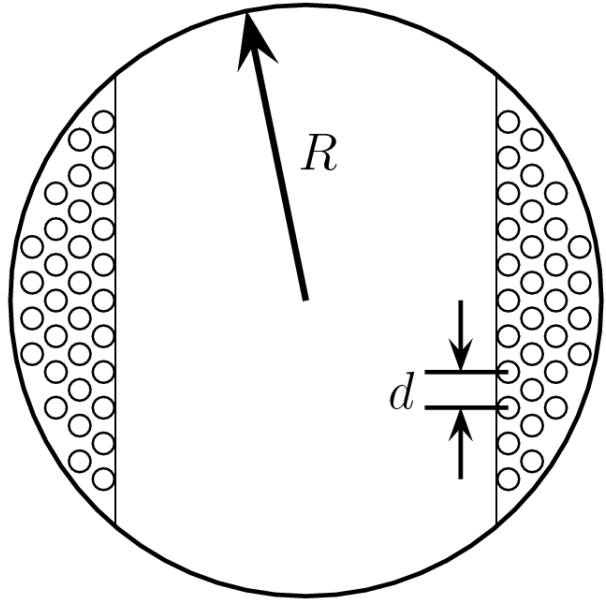}}
\caption{A model of bacteriophage genome packaging.
The viral capsid is modeled as a rigid 
spherical cavity. The DNA inside coils co-axially inward.
Neighboring DNA helices form a hexagonal lattice with lattice 
constant $d$. A sketch for a cross section of the viral capsid is shown. 
}
\label{fig:capsid}
\end{figure}

There exists different models to calculate the bending
energy of a packaged DNA molecules in literature 
[\onlinecite{Bloomfield78,Gelbart01,Phillips05,Harvey07,HarveyDNAcoil2008}].
In this paper, for simplicity, we employ the viral DNA packaging model 
used previously in Ref. \onlinecite{Phillips05}, \onlinecite{Bloomfield78},
\onlinecite{Gelbart01}.
In this model, the DNA viral genome are assumed to simply coil
co-axially inward with the neighboring DNA helices 
forming a hexagonal lattice with lattice constant $d$
(Fig. \ref{fig:capsid}). For a spherical capsid, this model gives:
\beq
E_{bend}(L_i,d) = \frac{4 \pi l_p k_B T}{\sqrt{3} d^2}
\Big\{-\Big(\frac{3\sqrt{3}L_id^2}{8\pi}\Big)^{1/3}
+ R \ln {
  \frac{R+({3\sqrt{3}L_id^2}/{8\pi})^{1/3}}
       {[ (R^2-({3\sqrt{3}L_id^2}/{8\pi})^{2/3} ]^{1/2}}
	    }
\Big\},
\label{Eq:Ebend}
\eeq
where $R$ is the radius of the inner surface of the viral capsid.

To calculate the interaction energy between neighboring
DNA segments inside the capsid, $E_{int}(L_i,d)$, we assume
that DNA molecules are almost neutralized by the 
counterions (the net charge, $\eta^*$ of the DNA segment
inside the capsid is much smaller than that of the ejected
segment because the latter has higher capacitance).
In the previous section, we have shown that for almost
neutral DNA, their interaction
is dominated by short range attraction forces.
Hence, one can approximate:
\beq
E_{int}(L_i,d_0) = -L_i |\mu_{DNA}|.
\label{Eq:EintSimple}
\eeq
Here, $d_0$ is the equilibrium interaxial distance of 
DNA bundle condensed by multivalent counterions.
Due to the strongly pressurized viral capsid, 
the actual interaxial distance, $d$, between neighboring DNA double helices
inside the capsid is smaller than the equilibrium distance, 
$d_0$, inside the condensate.
The experiments from Ref. \onlinecite{Parsegian92} provided an empirical
formula that relates the restoring force to the difference $d_0-d$.
Integrating this restoring force with $d$, one obtains 
an expression for the interaction energy between DNA helices
for a given interaxial distance $d$:
\beq
E_{int}(L_i,d)= L_i \sqrt{3} F_0 \Big[(c^2+c d) \exp{\Big(\frac{d_0-d}{c}\Big)}
-(c^2+c d_0) -\frac{1}{2}(d_0^2 - d^2)\Big]-L_i|\mu_{DNA}| ,
\label{Eq:Eint}
\eeq
where the empirical values of the constants $F_0$ and $c$ are
$0.5$ pN/nm$^2$ and $0.14$ nm respectively.

As we shown in the previous section, like the parameter $N_{Z,0}$, 
accurate calculation of $\mu_{DNA}$ is also very sensitive to 
an accurate determination of the counterion correlation energy, $\mu_{cor}$.
Adopting the same point of view, instead of using the 
analytical approximation Eq. (\ref{Eq:muAnalytical}),
we treat $\mu_{DNA}$ and $d_0$ as additional fitting parameters.
In total, our semi$-$empirical theory has three fitting parameters 
($N_{Z,0}$, $\mu_{DNA}$, $d_0$). From experimental data, we
have three fitting constrains (the two coordinates of the minimum
and the curvature of the curve $L_o(N_Z)$ in Fig. \ref{fig:Mg2}). 
Thus the theory does not contain unnecessary degrees of freedom.

\section{Fitting of experiment of DNA ejection from bacteriophages and discussion}

\label{sec:fitting}

Equation (\ref{Eq:Etotal}) together with equations 
(\ref{Eq:Eout}), (\ref{Eq:Ein}), (\ref{Eq:Ebend}) and 
(\ref{Eq:Eint}) provide the complete
expression for the total energy of the DNA genome of our 
semi-empirical theory.
For a given external osmotic pressure, $\Pi_{osm}$,
and a given multivalent counterion concentration, $N$, the equilibrium
value for the ejected DNA genome
length, $L_o^*$, is the length that minimizes the total free energy $G(L_o)$
of the system, where
\beq
G(L_o) = E_{tot}(L_o) + \Pi_{osm} L_o\pi a^2.
\label{Eq:Enthalpy}
\eeq
Here, $L_o\pi a^2$ is the volume of ejected DNA segments
in aqueous solution. 
The specific fitting procedure is following. 
The energy $E_{in}(L-L_o,d)$ of the DNA segment inside the capsid
is minimized with respect to $d$ to acquire
the optimal DNA-DNA interaxial distance for a given 
DNA ejected length, $d^*(L_o)$. 
Then, we substitute 
$E_{tot}(L_o) = E_{in}[L-L_o, d^*(L_o)] + E_{out}(L_o)$ into
Eq. (\ref{Eq:Enthalpy}) and optimize $G(L_o)$ with respect
to $L_o$ to obtain the equilibrium ejected length $L_o^*(\Pi_{osm},N)$.
By fitting $L_o^*$ with experiment data we can obtain the
values for the neutralizing counterion concentration, $N_{Z,0}$,
the Mg$^{+2}$ $-$ mediated DNA-DNA attraction, $-|\mu_{DNA}|$, and
the equilibrium DNA-DNA distance $d_0$.
The result of fitting our theoretical ejected length $L_o^*$ to
the experimental data of Ref. \onlinecite{Knobler08}
is shown in Fig. \ref{fig:Mg2}. In the
experiment, wild type bacteriophages $\lambda$ was used,
so $R=29$ nm and $L=16.49$ $\mu$m [\onlinecite{BakerLambdaCryoEM99}].
$\Pi_{osm}$
is held fixed at 3.5 atm and the Mg$^{+2}$ counterion concentration 
is varied from 10 mM to 200 mM. 
The fitted values are found to be $N_{Z,0}=64$ mM, 
$\mu_{DNA} = -0.004$ $k_BT$
per nucleotide base, and $d_0=2.73$ nm.

The strong influence of multivalent counterions on the process
of DNA ejection from bacteriophage appears in several aspects of our
theory and is easily seen by setting 
$d=d_0$, thus neglecting the weak dependence of $d$ on 
$L_i$ and using Eq. (\ref{Eq:EintSimple}) 
for DNA-DNA interactions inside the capsid. Firstly,
the attraction strength $|\mu_{DNA}|$ appears in the expression
for the free energy, Eq. (\ref{Eq:Enthalpy}), with the {\em same} sign
as $\Pi_{osm}$ (recall that $L_i = L - L_o$). In other words, the attraction between DNA strands
inside capsid acts as an additional ``effective" osmotic pressure
preventing the ejection of DNA from bacteriophage. 
This switch from repulsive DNA-DNA interactions for monovalent counterion
to attractive DNA-DNA interactions for Mg$^{+2}$
leads to an experimentally observed decrease in the percentage 
of DNA ejected from 50\% for monovalent counterions to 
20\% for Mg$^{+2}$ counterions at optimal inhibition ($N=N_{Z,0}$).
Secondly, the electrostatic energy of the ejected DNA segment
given by Eq. (\ref{Eq:Eout}) is logarithmically symmetrical around 
the neutralizing concentration $N_{Z,0}$. This is well demonstrated
in Fig. \ref{fig:Mg2} where the log-linear scale is used.
This symmetry is also similar to the behavior of another system
which exhibits charge inversion phenomenon, the non-monotonic
swelling of macroion by multivalent counterions [\onlinecite{Skinner08}].

It is very instructive to compare our fitting values for
$\mu_{DNA}$ and $N_{Z,0}$ 
to those obtained for other multivalent counterions.
Fitting done for the experiments of DNA condensation with
Spm$^{+4}$ and Spd$^{+3}$ shows 
$\mu_{DNA}$ to be $-$0.07 and $-$0.02 $k_BT$/base 
respectively [\onlinecite{Parsegian92}, \onlinecite{NguyenJCP2000}]. 
For our case of Mg$^{+2}$, a divalent counterion, 
and bacteriophage $\lambda$ experiment, 
$\mu_{DNA}$ is found to be $-0.004 k_BT$/base. This is quite
reasonable since Mg$^{+2}$ is a much weaker counterion
leading to much lower counterion correlation energy.
Furthermore, $N_{Z,0}$ was found to be
3.2 mM for the tetravalent counterion, 11 mM for
the trivalent counterion. Our fit of $N_{Z,0}=$64 mM for divalent
counterions again is in favorable agreement
with these independent fits. Note that in the limit
of high counterion valency ($Z\rightarrow\infty$), Eq. (\ref{Eq:c0analytical})
shows that $N_{Z,0}$ varies {\em exponentially} with $-Z^{3/2}$
[\onlinecite{Shklovskii1999,NguyenRMP2002,Netz02}].
The large increase in $N_{Z,0}$ from 3.2 mM for tetravalent
counterions to 11 mM for trivalent counterions,
and to 64mM for divalent counterions is not surprising.


It is quantitatively significant to point out
that our fitted value $\mu_{DNA}=-0.004k_BT$ per base 
explains why Mg$^{+2}$ ions cannot condense DNA in free solution.
This energy corresponds to 
an attraction of $-1.18k_BT$ per persistence length.
Since the thermal fluctuation energy of a polymer is about $k_BT$
per persistence length, this attraction is
too weak to overcome thermal fluctuations. It therefore
can only partially condense free DNA in solution [\onlinecite{Hud01}].
Only in the confinement
of the viral capsid can this attraction effect appear in 
the ejection process. It should be mentioned that computer simulations of
DNA condensation by idealized divalent counterions 
[\onlinecite{NguyenPRL2010}, \onlinecite{Nordenskiold95}]
show a weak short-range attraction comparable to our $\mu_{DNA}$.
The correlation induced DNA$-$DNA interaction obtained in the simulation
of Ref. \onlinecite{NguyenPRL2010} matches well with
our value of $-0.004k_BT$. This suggests that
in the presence of divalent counterions, electrostatic
interaction are an important (if not dominant) contribution to DNA$-$DNA 
short range interactions inside viral capsid. 

The phenomenological constants $\mu_{DNA} $ and $N_{Z,0}$ depend 
strongly on the strength of the correlations 
between multivalent counterions on the DNA surface. The stronger the
correlations, the greater the DNA$-$DNA attraction energy 
$|\mu_{DNA}|$ and the smaller
the concentration $N_{Z,0}$. In Ref. \onlinecite{Knobler08}, MgSO$_4$ salt induces
a strong inhibition effect. Due to this, $N_{Z,0}$ for MgSO$_4$ falls within 
the experimental measured concentration range
and we use these data to fit our theory. MgCl$_2$ induces weaker
inhibition, thus $N_{Z,0}$ for MgCl$_2$ is larger and apparently
lies at higher value than the measured range. More data
at higher MgCl$_2$ concentrations is needed to obtain
reliable fitting parameters for this case. 
In fact, the value $N_{Z,0} \simeq 104$ mM
obtained from the computer simulation of Ref. \onlinecite{NguyenPRL2010}
is nearly twice as large as our semi$-$empirical
results. This demonstrates again that this concentration is very
sensitive to the exact calculation of the counterion correlation
energy $\mu_{cor}$.
The authors of Ref. \onlinecite{Knobler08} also used non-ideality and 
ion specificity as an explanation for these differences. From our point
of view, they can lead to the difference in $\mu_{cor}$, hence in the value
$N_{Z,0}$. In the future, we plan to complimentary our phenomenological theory with a 
first principle calculation
to understand the ``microscopic" quantitative differences
between MgSO$_4$ and MgCl$_2$ salts. 

Lastly, we would like to point out that 
the fitted value for the equalibrium distance between neighboring
DNA in a bundle, $d_0 \simeq 27.3$\AA\ is well within the range
of various known distances from experiments 
[\onlinecite{Phillips05}, \onlinecite{Parsegian92}].

\section{Conclusion}

In conclusion,this paper has shown that divalent counterions such 
as Mg$^{+2}$ have strong effects on DNA condensation in a confined 
environment (such as inside bacteriophages capsid) 
similar to those of counterions with higher valency. 
We propose that the non-monotonic dependence of the amount
of DNA ejected from bacteriophages has the same physical
origin as the reentrant condensation phenomenon of DNA
molecules by multivalent counterions. Fitting our semi-empirical
theory to available experimental data, we obtain the strength
of DNA$-$DNA short range attraction mediated by divalent
counterions. The fitted values agree quantitatively 
and qualitatively with experimental values from other
DNA system and computer simulations. This shows that
in the problem of viral DNA package where DNA lateral
motion is restricted, divalent counterions can plays an important
role similar to that of counterions with higher valency.
This fact should to be incorporated in any electrostatic theories of bacteriophage
packaging. The strength of short-range DNA-DNA attractions 
mediated by MgSO$_4$ salt is first obtained by the authors. 
It provides a good starting point for future works with DNA-DNA condensation
in the presence of divalent counterions.


\begin{acknowledgments}
We would like to thank Doctors Shklovskii, Evilevitch, Fang, Gelbart, Podgornik, Naji,
Phillips, Rau, and Parsegian for valuable discussions. TTN acknowledges the
hospitality of the Fine Theoretical Physics Institute and the Aspen Physics
Center where part of this work was done. TTN acknowledges 
the support of junior faculty from the Georgia Institute of Technology.
SL acknowledges financial support from Korean-American Scientists
and Engineers Association (Georgia chapter).
\end{acknowledgments}

\bibliography{nttpaper}

\begin{thebibliography}{31}%
\makeatletter
\providecommand \@ifxundefined [1]{%
 \@ifx{#1\undefined}
}%
\providecommand \@ifnum [1]{%
 \ifnum #1\expandafter \@firstoftwo
 \else \expandafter \@secondoftwo
 \fi
}%
\providecommand \@ifx [1]{%
 \ifx #1\expandafter \@firstoftwo
 \else \expandafter \@secondoftwo
 \fi
}%
\providecommand \natexlab [1]{#1}%
\providecommand \enquote  [1]{``#1''}%
\providecommand \bibnamefont  [1]{#1}%
\providecommand \bibfnamefont [1]{#1}%
\providecommand \citenamefont [1]{#1}%
\providecommand \href@noop [0]{\@secondoftwo}%
\providecommand \href [0]{\begingroup \@sanitize@url \@href}%
\providecommand \@href[1]{\@@startlink{#1}\@@href}%
\providecommand \@@href[1]{\endgroup#1\@@endlink}%
\providecommand \@sanitize@url [0]{\catcode `\\12\catcode `\$12\catcode
  `\&12\catcode `\#12\catcode `\^12\catcode `\_12\catcode `\%12\relax}%
\providecommand \@@startlink[1]{}%
\providecommand \@@endlink[0]{}%
\providecommand \url  [0]{\begingroup\@sanitize@url \@url }%
\providecommand \@url [1]{\endgroup\@href {#1}{\urlprefix }}%
\providecommand \urlprefix  [0]{URL }%
\providecommand \Eprint [0]{\href }%
\providecommand \doibase [0]{http://dx.doi.org/}%
\providecommand \selectlanguage [0]{\@gobble}%
\providecommand \bibinfo  [0]{\@secondoftwo}%
\providecommand \bibfield  [0]{\@secondoftwo}%
\providecommand \translation [1]{[#1]}%
\providecommand \BibitemOpen [0]{}%
\providecommand \bibitemStop [0]{}%
\providecommand \bibitemNoStop [0]{.\EOS\space}%
\providecommand \EOS [0]{\spacefactor3000\relax}%
\providecommand \BibitemShut  [1]{\csname bibitem#1\endcsname}%
\let\auto@bib@innerbib\@empty
\bibitem [{\citenamefont {Smith}\ \emph {et~al.}(2001)\citenamefont {Smith},
  \citenamefont {Trans}, \citenamefont {Smith}, \citenamefont {Grimes},
  \citenamefont {Anderson},\ and\ \citenamefont {Bustamante}}]{Bustamante01}%
  \BibitemOpen
  \bibfield  {author} {\bibinfo {author} {\bibfnamefont {D.~E.}\ \bibnamefont
  {Smith}}, \bibinfo {author} {\bibfnamefont {S.~J.}\ \bibnamefont {Trans}},
  \bibinfo {author} {\bibfnamefont {S.~B.}\ \bibnamefont {Smith}}, \bibinfo
  {author} {\bibfnamefont {S.}~\bibnamefont {Grimes}}, \bibinfo {author}
  {\bibfnamefont {D.~L.}\ \bibnamefont {Anderson}}, \ and\ \bibinfo {author}
  {\bibfnamefont {C.}~\bibnamefont {Bustamante}},\ }\href@noop {} {\bibfield
  {journal} {\bibinfo  {journal} {Nature}\ }\textbf {\bibinfo {volume} {413}},\
  \bibinfo {pages} {748} (\bibinfo {year} {2001})}\BibitemShut {NoStop}%
\bibitem [{\citenamefont {Evilevitch}\ \emph {et~al.}(2003)\citenamefont
  {Evilevitch}, \citenamefont {Lavelle}, \citenamefont {Knobler}, \citenamefont
  {Raspaud},\ and\ \citenamefont {Gelbart}}]{Gelbart03}%
  \BibitemOpen
  \bibfield  {author} {\bibinfo {author} {\bibfnamefont {A.}~\bibnamefont
  {Evilevitch}}, \bibinfo {author} {\bibfnamefont {L.}~\bibnamefont {Lavelle}},
  \bibinfo {author} {\bibfnamefont {C.~M.}\ \bibnamefont {Knobler}}, \bibinfo
  {author} {\bibfnamefont {E.}~\bibnamefont {Raspaud}}, \ and\ \bibinfo
  {author} {\bibfnamefont {W.~M.}\ \bibnamefont {Gelbart}},\ }\href@noop {}
  {\bibfield  {journal} {\bibinfo  {journal} {Proc. Nat. Acad. Sci. USA}\
  }\textbf {\bibinfo {volume} {100}},\ \bibinfo {pages} {9292} (\bibinfo {year}
  {2003})}\BibitemShut {NoStop}%
\bibitem [{\citenamefont {Castelnovo}\ \emph {et~al.}(2003)\citenamefont
  {Castelnovo}, \citenamefont {Bowles}, \citenamefont {Reiss},\ and\
  \citenamefont {Gelbart}}]{Gelbart2003}%
  \BibitemOpen
  \bibfield  {author} {\bibinfo {author} {\bibfnamefont {M.}~\bibnamefont
  {Castelnovo}}, \bibinfo {author} {\bibfnamefont {R.~K.}\ \bibnamefont
  {Bowles}}, \bibinfo {author} {\bibfnamefont {H.}~\bibnamefont {Reiss}}, \
  and\ \bibinfo {author} {\bibfnamefont {W.~M.}\ \bibnamefont {Gelbart}},\
  }\href@noop {} {\bibfield  {journal} {\bibinfo  {journal} {Eur. Phys. J. E}\
  }\textbf {\bibinfo {volume} {10}},\ \bibinfo {pages} {191} (\bibinfo {year}
  {2003})}\BibitemShut {NoStop}%
\bibitem [{\citenamefont {Petrov}, \citenamefont {Lim-Hing},\ and\
  \citenamefont {Harvey}(2007)}]{Harvey07}%
  \BibitemOpen
  \bibfield  {author} {\bibinfo {author} {\bibfnamefont {A.~S.}\ \bibnamefont
  {Petrov}}, \bibinfo {author} {\bibfnamefont {K.}~\bibnamefont {Lim-Hing}}, \
  and\ \bibinfo {author} {\bibfnamefont {S.~C.}\ \bibnamefont {Harvey}},\
  }\href@noop {} {\bibfield  {journal} {\bibinfo  {journal} {Structure}\
  }\textbf {\bibinfo {volume} {15}},\ \bibinfo {pages} {807} (\bibinfo {year}
  {2007})}\BibitemShut {NoStop}%
\bibitem [{\citenamefont {Letellier}\ \emph {et~al.}(2004)\citenamefont
  {Letellier}, \citenamefont {Boulanger}, \citenamefont {Plancon},
  \citenamefont {Jacquot},\ and\ \citenamefont {Santamaria}}]{Santamaria04}%
  \BibitemOpen
  \bibfield  {author} {\bibinfo {author} {\bibfnamefont {L.}~\bibnamefont
  {Letellier}}, \bibinfo {author} {\bibfnamefont {P.}~\bibnamefont
  {Boulanger}}, \bibinfo {author} {\bibfnamefont {L.}~\bibnamefont {Plancon}},
  \bibinfo {author} {\bibfnamefont {P.}~\bibnamefont {Jacquot}}, \ and\
  \bibinfo {author} {\bibfnamefont {M.}~\bibnamefont {Santamaria}},\
  }\href@noop {} {\bibfield  {journal} {\bibinfo  {journal} {Front. Biosci.}\
  }\textbf {\bibinfo {volume} {9}},\ \bibinfo {pages} {1228} (\bibinfo {year}
  {2004})}\BibitemShut {NoStop}%
\bibitem [{\citenamefont {Black}(1989)}]{Black89}%
  \BibitemOpen
  \bibfield  {author} {\bibinfo {author} {\bibfnamefont {L.~W.}\ \bibnamefont
  {Black}},\ }\href@noop {} {\bibfield  {journal} {\bibinfo  {journal} {Annu.
  Rev. Microbiol.}\ }\textbf {\bibinfo {volume} {43}},\ \bibinfo {pages} {267}
  (\bibinfo {year} {1989})}\BibitemShut {NoStop}%
\bibitem [{\citenamefont {Murialdo}(1991)}]{Murialdo91}%
  \BibitemOpen
  \bibfield  {author} {\bibinfo {author} {\bibfnamefont {H.}~\bibnamefont
  {Murialdo}},\ }\href@noop {} {\bibfield  {journal} {\bibinfo  {journal}
  {Annu. Rev. Biochem.}\ }\textbf {\bibinfo {volume} {60}},\ \bibinfo {pages}
  {125} (\bibinfo {year} {1991})}\BibitemShut {NoStop}%
\bibitem [{\citenamefont {Purohit}\ \emph {et~al.}(2005)\citenamefont
  {Purohit}, \citenamefont {Inamdar}, \citenamefont {Grayson}, \citenamefont
  {Squires}, \citenamefont {Kondev},\ and\ \citenamefont
  {Phillips}}]{Phillips05}%
  \BibitemOpen
  \bibfield  {author} {\bibinfo {author} {\bibfnamefont {P.~K.}\ \bibnamefont
  {Purohit}}, \bibinfo {author} {\bibfnamefont {M.~M.}\ \bibnamefont
  {Inamdar}}, \bibinfo {author} {\bibfnamefont {P.~D.}\ \bibnamefont
  {Grayson}}, \bibinfo {author} {\bibfnamefont {T.~M.}\ \bibnamefont
  {Squires}}, \bibinfo {author} {\bibfnamefont {J.}~\bibnamefont {Kondev}}, \
  and\ \bibinfo {author} {\bibfnamefont {R.}~\bibnamefont {Phillips}},\
  }\href@noop {} {\bibfield  {journal} {\bibinfo  {journal} {Biophys. J.}\
  }\textbf {\bibinfo {volume} {88}},\ \bibinfo {pages} {851} (\bibinfo {year}
  {2005})}\BibitemShut {NoStop}%
\bibitem [{\citenamefont {Evilevitch}\ \emph {et~al.}(2004)\citenamefont
  {Evilevitch}, \citenamefont {Castelnovo}, \citenamefont {Knobler},\ and\
  \citenamefont {Gelbart}}]{Gelbart04}%
  \BibitemOpen
  \bibfield  {author} {\bibinfo {author} {\bibfnamefont {A.}~\bibnamefont
  {Evilevitch}}, \bibinfo {author} {\bibfnamefont {M.}~\bibnamefont
  {Castelnovo}}, \bibinfo {author} {\bibfnamefont {C.~M.}\ \bibnamefont
  {Knobler}}, \ and\ \bibinfo {author} {\bibfnamefont {W.~M.}\ \bibnamefont
  {Gelbart}},\ }\href@noop {} {\bibfield  {journal} {\bibinfo  {journal} {J.
  Phys. Chem. B}\ }\textbf {\bibinfo {volume} {108}},\ \bibinfo {pages} {6838}
  (\bibinfo {year} {2004})}\BibitemShut {NoStop}%
\bibitem [{\citenamefont {Evilevitch}\ \emph {et~al.}(2008)\citenamefont
  {Evilevitch}, \citenamefont {Fang}, \citenamefont {Yoffe}, \citenamefont
  {Castelnovo}, \citenamefont {Rau}, \citenamefont {Parsegian}, \citenamefont
  {Gelbart},\ and\ \citenamefont {Knobler}}]{Knobler08}%
  \BibitemOpen
  \bibfield  {author} {\bibinfo {author} {\bibfnamefont {A.}~\bibnamefont
  {Evilevitch}}, \bibinfo {author} {\bibfnamefont {L.~T.}\ \bibnamefont
  {Fang}}, \bibinfo {author} {\bibfnamefont {A.~M.}\ \bibnamefont {Yoffe}},
  \bibinfo {author} {\bibfnamefont {M.}~\bibnamefont {Castelnovo}}, \bibinfo
  {author} {\bibfnamefont {D.~C.}\ \bibnamefont {Rau}}, \bibinfo {author}
  {\bibfnamefont {V.~A.}\ \bibnamefont {Parsegian}}, \bibinfo {author}
  {\bibfnamefont {W.~M.}\ \bibnamefont {Gelbart}}, \ and\ \bibinfo {author}
  {\bibfnamefont {C.~M.}\ \bibnamefont {Knobler}},\ }\href@noop {} {\bibfield
  {journal} {\bibinfo  {journal} {Biophys. J.}\ }\textbf {\bibinfo {volume}
  {94}},\ \bibinfo {pages} {1110} (\bibinfo {year} {2008})}\BibitemShut
  {NoStop}%
\bibitem [{\citenamefont {Fang}(2009)}]{GelbartPriCom09}%
  \BibitemOpen
  \bibfield  {author} {\bibinfo {author} {\bibfnamefont {L.~T.}\ \bibnamefont
  {Fang}},\ }\href@noop {} {}\bibinfo {howpublished} {private communications}
  (\bibinfo {year} {2009})\BibitemShut {NoStop}%
\bibitem [{\citenamefont {Shklovskii}(1999)}]{Shklovskii1999}%
  \BibitemOpen
  \bibfield  {author} {\bibinfo {author} {\bibfnamefont {B.~I.}\ \bibnamefont
  {Shklovskii}},\ }\href@noop {} {\bibfield  {journal} {\bibinfo  {journal}
  {Phys. Rev. E}\ }\textbf {\bibinfo {volume} {60}},\ \bibinfo {pages} {5802}
  (\bibinfo {year} {1999})}\BibitemShut {NoStop}%
\bibitem [{\citenamefont {Grosberg}, \citenamefont {Nguyen},\ and\
  \citenamefont {Shklovskii}(2002)}]{NguyenRMP2002}%
  \BibitemOpen
  \bibfield  {author} {\bibinfo {author} {\bibfnamefont {A.~Y.}\ \bibnamefont
  {Grosberg}}, \bibinfo {author} {\bibfnamefont {T.~T.}\ \bibnamefont
  {Nguyen}}, \ and\ \bibinfo {author} {\bibfnamefont {B.}~\bibnamefont
  {Shklovskii}},\ }\href@noop {} {\bibfield  {journal} {\bibinfo  {journal}
  {Rev. Mod. Phys.}\ }\textbf {\bibinfo {volume} {74}},\ \bibinfo {pages} {329}
  (\bibinfo {year} {2002})}\BibitemShut {NoStop}%
\bibitem [{\citenamefont {Moreira}\ and\ \citenamefont {Netz}(2002)}]{Netz02}%
  \BibitemOpen
  \bibfield  {author} {\bibinfo {author} {\bibfnamefont {A.~G.}\ \bibnamefont
  {Moreira}}\ and\ \bibinfo {author} {\bibfnamefont {R.~R.}\ \bibnamefont
  {Netz}},\ }\href@noop {} {\bibfield  {journal} {\bibinfo  {journal} {Eur.
  Phys. J. E}\ }\textbf {\bibinfo {volume} {8}},\ \bibinfo {pages} {33}
  (\bibinfo {year} {2002})}\BibitemShut {NoStop}%
\bibitem [{\citenamefont {Besteman}, \citenamefont {Eijk},\ and\ \citenamefont
  {Lemay}(2007)}]{lemayNature2007}%
  \BibitemOpen
  \bibfield  {author} {\bibinfo {author} {\bibfnamefont {K.}~\bibnamefont
  {Besteman}}, \bibinfo {author} {\bibfnamefont {K.~V.}\ \bibnamefont {Eijk}},
  \ and\ \bibinfo {author} {\bibfnamefont {S.~G.}\ \bibnamefont {Lemay}},\
  }\href@noop {} {\bibfield  {journal} {\bibinfo  {journal} {Nature Physics}\
  }\textbf {\bibinfo {volume} {3}},\ \bibinfo {pages} {641} (\bibinfo {year}
  {2007})}\BibitemShut {NoStop}%
\bibitem [{\citenamefont {Kandu{\v{c}}}, \citenamefont {Naji},\ and\
  \citenamefont {Podgornik}(2010)}]{KanduJCP2010}%
  \BibitemOpen
  \bibfield  {author} {\bibinfo {author} {\bibfnamefont {M.}~\bibnamefont
  {Kandu{\v{c}}}}, \bibinfo {author} {\bibfnamefont {A.}~\bibnamefont {Naji}},
  \ and\ \bibinfo {author} {\bibfnamefont {R.}~\bibnamefont {Podgornik}},\
  }\href@noop {} {\bibfield  {journal} {\bibinfo  {journal} {J. Chem. Phys.}\
  }\textbf {\bibinfo {volume} {132}},\ \bibinfo {pages} {224703} (\bibinfo
  {year} {2010})}\BibitemShut {NoStop}%
\bibitem [{\citenamefont {Rau}\ and\ \citenamefont
  {Parsegian}(1992)}]{Parsegian92}%
  \BibitemOpen
  \bibfield  {author} {\bibinfo {author} {\bibfnamefont {D.~C.}\ \bibnamefont
  {Rau}}\ and\ \bibinfo {author} {\bibfnamefont {V.~A.}\ \bibnamefont
  {Parsegian}},\ }\href@noop {} {\bibfield  {journal} {\bibinfo  {journal}
  {Biophys. J.}\ }\textbf {\bibinfo {volume} {61}},\ \bibinfo {pages} {246}
  (\bibinfo {year} {1992})}\BibitemShut {NoStop}%
\bibitem [{\citenamefont {Hud}\ and\ \citenamefont {Downing}(2001)}]{Hud01}%
  \BibitemOpen
  \bibfield  {author} {\bibinfo {author} {\bibfnamefont {N.~V.}\ \bibnamefont
  {Hud}}\ and\ \bibinfo {author} {\bibfnamefont {K.~H.}\ \bibnamefont
  {Downing}},\ }\href@noop {} {\bibfield  {journal} {\bibinfo  {journal} {Proc.
  Nat. Acad. Sci. USA}\ }\textbf {\bibinfo {volume} {98}},\ \bibinfo {pages}
  {14925} (\bibinfo {year} {2001})}\BibitemShut {NoStop}%
\bibitem [{\citenamefont {Koltover}, \citenamefont {Wagner},\ and\
  \citenamefont {Safinya}(2000)}]{Koltover2000}%
  \BibitemOpen
  \bibfield  {author} {\bibinfo {author} {\bibfnamefont {I.}~\bibnamefont
  {Koltover}}, \bibinfo {author} {\bibfnamefont {K.}~\bibnamefont {Wagner}}, \
  and\ \bibinfo {author} {\bibfnamefont {C.~R.}\ \bibnamefont {Safinya}},\
  }\href@noop {} {\bibfield  {journal} {\bibinfo  {journal} {Proc. Nat. Acad.
  Sci. USA}\ }\textbf {\bibinfo {volume} {97}},\ \bibinfo {pages} {14046}
  (\bibinfo {year} {2000})}\BibitemShut {NoStop}%
\bibitem [{\citenamefont {Lee}, \citenamefont {Le},\ and\ \citenamefont
  {Nguyen}(2010)}]{NguyenPRL2010}%
  \BibitemOpen
  \bibfield  {author} {\bibinfo {author} {\bibfnamefont {S.}~\bibnamefont
  {Lee}}, \bibinfo {author} {\bibfnamefont {T.~T.}\ \bibnamefont {Le}}, \ and\
  \bibinfo {author} {\bibfnamefont {T.~T.}\ \bibnamefont {Nguyen}},\
  }\href@noop {} {\bibfield  {journal} {\bibinfo  {journal} {Phys. Rev. Lett.
  (Accepted for publication)}\ } (\bibinfo {year} {2010})},\ \Eprint
  {http://arxiv.org/abs/cond-mat/0912.3595} {arXiv:cond-mat/0912.3595}
  \BibitemShut {NoStop}%
\bibitem [{\citenamefont {Lyubartsev}\ and\ \citenamefont
  {Nordenski{\"{o}}ld}(1995)}]{Nordenskiold95}%
  \BibitemOpen
  \bibfield  {author} {\bibinfo {author} {\bibfnamefont {A.~P.}\ \bibnamefont
  {Lyubartsev}}\ and\ \bibinfo {author} {\bibfnamefont {L.}~\bibnamefont
  {Nordenski{\"{o}}ld}},\ }\href@noop {} {\bibfield  {journal} {\bibinfo
  {journal} {J. Phys. Chem.}\ }\textbf {\bibinfo {volume} {99}},\ \bibinfo
  {pages} {10373} (\bibinfo {year} {1995})}\BibitemShut {NoStop}%
\bibitem [{\citenamefont {Winterhalter}\ and\ \citenamefont
  {Helfrich}(1988)}]{Winterhalter88}%
  \BibitemOpen
  \bibfield  {author} {\bibinfo {author} {\bibfnamefont {M.}~\bibnamefont
  {Winterhalter}}\ and\ \bibinfo {author} {\bibfnamefont {W.}~\bibnamefont
  {Helfrich}},\ }\href@noop {} {\bibfield  {journal} {\bibinfo  {journal} {J.
  Phys. Chem.}\ }\textbf {\bibinfo {volume} {92}},\ \bibinfo {pages} {6865}
  (\bibinfo {year} {1988})}\BibitemShut {NoStop}%
\bibitem [{\citenamefont {Perel}\ and\ \citenamefont
  {Shklovskii}(1999)}]{Shklovskii99}%
  \BibitemOpen
  \bibfield  {author} {\bibinfo {author} {\bibfnamefont {V.~I.}\ \bibnamefont
  {Perel}}\ and\ \bibinfo {author} {\bibfnamefont {B.~I.}\ \bibnamefont
  {Shklovskii}},\ }\href@noop {} {\bibfield  {journal} {\bibinfo  {journal}
  {Physica A}\ }\textbf {\bibinfo {volume} {274}},\ \bibinfo {pages} {446}
  (\bibinfo {year} {1999})}\BibitemShut {NoStop}%
\bibitem [{\citenamefont {Bruinsma}(1998)}]{BruinsmaCounterionRelease1998}%
  \BibitemOpen
  \bibfield  {author} {\bibinfo {author} {\bibfnamefont {R.}~\bibnamefont
  {Bruinsma}},\ }\href@noop {} {\bibfield  {journal} {\bibinfo  {journal} {Eur.
  Phys. J. B}\ }\textbf {\bibinfo {volume} {4}},\ \bibinfo {pages} {75}
  (\bibinfo {year} {1998})}\BibitemShut {NoStop}%
\bibitem [{\citenamefont {Gelbart}\ \emph {et~al.}(2000)\citenamefont
  {Gelbart}, \citenamefont {Bruinsma}, \citenamefont {Pincus},\ and\
  \citenamefont {Parsegian}}]{GelbartPhysToday}%
  \BibitemOpen
  \bibfield  {author} {\bibinfo {author} {\bibfnamefont {W.~M.}\ \bibnamefont
  {Gelbart}}, \bibinfo {author} {\bibfnamefont {R.~F.}\ \bibnamefont
  {Bruinsma}}, \bibinfo {author} {\bibfnamefont {P.~A.}\ \bibnamefont
  {Pincus}}, \ and\ \bibinfo {author} {\bibfnamefont {A.~V.}\ \bibnamefont
  {Parsegian}},\ }\href@noop {} {\bibfield  {journal} {\bibinfo  {journal}
  {Phys. Today}\ }\textbf {\bibinfo {volume} {53}},\ \bibinfo {pages} {38}
  (\bibinfo {year} {2000})}\BibitemShut {NoStop}%
\bibitem [{\citenamefont {Riemer}\ and\ \citenamefont
  {Bloomfield}(1978)}]{Bloomfield78}%
  \BibitemOpen
  \bibfield  {author} {\bibinfo {author} {\bibfnamefont {S.~C.}\ \bibnamefont
  {Riemer}}\ and\ \bibinfo {author} {\bibfnamefont {V.~A.}\ \bibnamefont
  {Bloomfield}},\ }\href@noop {} {\bibfield  {journal} {\bibinfo  {journal}
  {Biopolymers}\ }\textbf {\bibinfo {volume} {17}},\ \bibinfo {pages} {785}
  (\bibinfo {year} {1978})}\BibitemShut {NoStop}%
\bibitem [{\citenamefont {Kindt}\ \emph {et~al.}(2001)\citenamefont {Kindt},
  \citenamefont {Tzlil}, \citenamefont {Ben-Shaul},\ and\ \citenamefont
  {Gelbart}}]{Gelbart01}%
  \BibitemOpen
  \bibfield  {author} {\bibinfo {author} {\bibfnamefont {J.}~\bibnamefont
  {Kindt}}, \bibinfo {author} {\bibfnamefont {S.}~\bibnamefont {Tzlil}},
  \bibinfo {author} {\bibfnamefont {A.}~\bibnamefont {Ben-Shaul}}, \ and\
  \bibinfo {author} {\bibfnamefont {W.~M.}\ \bibnamefont {Gelbart}},\
  }\href@noop {} {\bibfield  {journal} {\bibinfo  {journal} {Proc. Nat. Acad.
  Sci. USA}\ }\textbf {\bibinfo {volume} {98}},\ \bibinfo {pages} {13671}
  (\bibinfo {year} {2001})}\BibitemShut {NoStop}%
\bibitem [{\citenamefont {Petrov}\ and\ \citenamefont
  {Harvey}(2008)}]{HarveyDNAcoil2008}%
  \BibitemOpen
  \bibfield  {author} {\bibinfo {author} {\bibfnamefont {A.~S.}\ \bibnamefont
  {Petrov}}\ and\ \bibinfo {author} {\bibfnamefont {S.~C.}\ \bibnamefont
  {Harvey}},\ }\href@noop {} {\bibfield  {journal} {\bibinfo  {journal}
  {Biophys. J.}\ }\textbf {\bibinfo {volume} {95}},\ \bibinfo {pages} {497}
  (\bibinfo {year} {2008})}\BibitemShut {NoStop}%
\bibitem [{\citenamefont {Baker}, \citenamefont {Olson},\ and\ \citenamefont
  {Fuller}(1999)}]{BakerLambdaCryoEM99}%
  \BibitemOpen
  \bibfield  {author} {\bibinfo {author} {\bibfnamefont {T.~S.}\ \bibnamefont
  {Baker}}, \bibinfo {author} {\bibfnamefont {N.~H.}\ \bibnamefont {Olson}}, \
  and\ \bibinfo {author} {\bibfnamefont {S.~D.}\ \bibnamefont {Fuller}},\
  }\href@noop {} {\bibfield  {journal} {\bibinfo  {journal} {Microbiol. Mol.
  Biol. Rev.}\ }\textbf {\bibinfo {volume} {63}},\ \bibinfo {pages} {862}
  (\bibinfo {year} {1999})}\BibitemShut {NoStop}%
\bibitem [{\citenamefont {Skinner}\ and\ \citenamefont
  {Shklovskii}(2009)}]{Skinner08}%
  \BibitemOpen
  \bibfield  {author} {\bibinfo {author} {\bibfnamefont {B.}~\bibnamefont
  {Skinner}}\ and\ \bibinfo {author} {\bibfnamefont {B.~I.}\ \bibnamefont
  {Shklovskii}},\ }\href@noop {} {\bibfield  {journal} {\bibinfo  {journal}
  {Physica A}\ }\textbf {\bibinfo {volume} {388}},\ \bibinfo {pages} {1}
  (\bibinfo {year} {2009})}\BibitemShut {NoStop}%
\bibitem [{\citenamefont {Nguyen}, \citenamefont {Rouzina},\ and\ \citenamefont
  {Shklovskii}(2000)}]{NguyenJCP2000}%
  \BibitemOpen
  \bibfield  {author} {\bibinfo {author} {\bibfnamefont {T.~T.}\ \bibnamefont
  {Nguyen}}, \bibinfo {author} {\bibfnamefont {I.}~\bibnamefont {Rouzina}}, \
  and\ \bibinfo {author} {\bibfnamefont {B.~I.}\ \bibnamefont {Shklovskii}},\
  }\href@noop {} {\bibfield  {journal} {\bibinfo  {journal} {J. Chem. Phys.}\
  }\textbf {\bibinfo {volume} {112}},\ \bibinfo {pages} {2562} (\bibinfo {year}
  {2000})}\BibitemShut {NoStop}%
\end{thebibliography}%

\end{document}